\documentclass[twocolumn,groupedaddress,assymb,amsmath]{revtex4}
\usepackage{graphicx}
\usepackage{amssymb}
\usepackage{amsmath}

\usepackage{extarrows}
\usepackage[dvipsnames]{xcolor}
\usepackage[colorlinks,citecolor=blue, linkcolor=blue,hyperindex,CJKbookmarks,dvipdfm]{hyperref}
\usepackage{lipsum} 

\usepackage{multirow} 
\begin{document}
\title{Exponentially enhanced gravitationally induced entanglement between quantum systems with a two-phonon drive}
  \author{Dianzhen Cui$^{1}$, X. X. Yi$^{1,2}$\footnote{yixx@nenu.edu.cn}}
\affiliation{$^1$Center for Quantum Sciences and School of Physics, Northeast Normal University, Changchun 130024, China\\
$^2$Center for Advanced Optoelectronic Functional Materials Research, and Key Laboratory for UV Light-Emitting
{Materials and Technology of Ministry of Education, Northeast Normal University, Changchun 130024, China}}

  \date{\today}

 \begin{abstract}
Finding a feasible protocol for probing the quantum nature of gravity has been attracting an increasing amount of attention. In this manuscript, we propose a protocol to enhance the detection of gravitationally induced  entanglement by exploiting the two-phonon drive in a hybrid quantum setup. We consider the setup consisting of a test particle in a double-well potential, a qubit and a quantum mediator. There is gravitational interaction between the test particle and the mediator, and a spin-phonon coupling between the mediator and the qubit. By introducing a two-phonon drive, the  entanglement between the TP and the qubit are significantly enhanced and the entanglement generation rate is remarkably increased  compared with the case without the two-phonon drive. Moreover, the entanglement between the TP and the qubit can be partially preserved in the presence of dephasing by the proposed strategy. This work would open a different avenue for experimental detection of the quantum nature of gravity, which could find applications in quantum information science.
 \end{abstract}
\maketitle

\section{Introduction}
Sixty years ago, physicists gathered to discuss gravity at the Chapel Hill conference \cite{Rickles2011}. Richard Feynman proposed a thought experiment to analyze a profound problem: the incompatibility of general relativity and quantum theory. First, he considered  a mass is prepared in a quantum superposition of two different locations, $A$ and $B$. General relativity describes how an object interacts with the gravitational field: the mass drops depending on the strength of gravity locally and also slightly changes the value of the field at $A$ and $B$
by its presence. Applying both theories means that the gravitational field must also assume two configurations at once: corresponding to the case where the mass is at $A$ or $B$. In other words, gravity would take on a nonclassical nature when it interacts with a mass that behaves in a quantum way.

Although the thought experiment is interesting, one main difficulty to judge to incompatibility  is the lack of experimental evidence. Fortunately, recent advances in the field of optomechanics have provided the possibility to detect  the quantum nature of gravity \cite{Aspelmeyer2014,OConnell2010,Chan2011,
Delic2020,Magrini2021,Matsumoto2020,Quidant2021,Pedernales2021,Whittle2021}, making the judgement of the incompatibility feasible. Marletto $et$ $al$. \cite{Marletto2017witness} proposed an interference protocol to witness quantum nature of the gravitational field by detecting it with two masses. The first mass, being in a quantum superposition of two locations, produces Schr\"{o}dinger-cat states on the gravitational field through gravitational interaction. The second mass (the quantum probe) then interacts with the field and is measured to reveal the quantum nature of the field. Such an protocol requires an extra step than in Feynman's original thought experiment.  Krisnanda $et$ $al$. \cite{Krisnanda2017} proposed a scheme for revealing the nonclassicality of the gravitational field characterized by quantum discord. The entanglement generation between two gravitationally interacting systems   signifies the nonclassicality of the gravitational field between them. This has motivated experimental proposals to explore the gravity-induced entanglement \cite{Bose2017,Marletto2017,Qvarfort2019,Miao2020,Paterek2020,Weiss2021}. In the literature, proposed schemes typically rely on the direct gravitational interaction between two heavy masses. Most recently \cite{Plenio2022}, a protocol is proposed to enhance this gravitational interaction by utilizing a massive particle as a mediator to induce  the interaction between a two-level test mass (system $a$) and an ancillary qubit (system $b$). The mediator (system $c$) is weakly coupled to the test mass through gravitational interaction with a coupling strength of $g_a$ and strongly coupled to the ancillary qubit with coupling strength $g_b\gg g_a$. At appropriately chosen points in time, the mediator decorrelates from the system, leaving two light masses entangled [with the unitary-evolution operator $U(t_n)=\exp(-ig_{\text{eff}} \sigma_a^z\sigma_b^z t_n)$, $g_{\text{eff}}$ is the effective interaction between two light masses]. An effective interaction between two light masses that increases with the mass of the mediator is achieved, which is independent of its initial state. While this is a truly remarkable development, the experimental implementation of such entanglement is still extremely challenging given the difficulty in achieving quantum control of mechanical degrees of freedom of ever more massive particles \cite{Aspelmeyer2016}. On the other hand, all physical systems are invariably dissipative as they  inevitably couple  to their environment \cite{Schlosshauer2007,Mazumdar2022,Matsumura2022Path,Matsumura2022,Miki2021,
Cui2021,Bose2021Relative,Cui2022,Yi2022,Budini00400,Eberly2002,
Eberly2003,Clerk2022,Kampen1992,Tarn2009,Reina2002,Schreier2008,Clerk2008,Delord2017}. As a consequence, the macroscopic manifestation of this quantum effect is extremely fragile and is dramatically reduced due to the decoherence of the systems. This makes the observation of  such an entanglement challenging. Therefore, developing new techniques for detecting the quantum nature of gravity is highly desirable.

A feasible way may lie in using a squeezing protocol to realize non-local controllable interaction between quantum systems.
In this manuscript, we theoretically propose a mechanism to enhance the amount of  gravitationally induced  entanglement by exploiting the mechanical squeezing in a hybrid quantum system. The proposed system contains a test particle (TP) trapped in a double-well potential (DWP) (manifested as a two-level system), a qubit, and a mechanical oscillator (mediator). The mediator is weakly coupled to the TP through gravitational interaction and strongly coupled to the qubit through a spin-phonon magnetic
coupling. The qubit-mediator coupling strength that does not depend on the mass of the qubit. Therefore, our scheme has only two masses [TP ($m_{\tilde{a}}$) and mediator ($m_{\tilde{c}}$)]. Since the TP-mediator and qubit-mediator interactions depend linearly on the position operator of the mediator, we can enhance the interaction by increasing the spatial extent of the center-of-mass (COM) wave function using a squeezing protocol. This squeezing is implemented through the Coulomb interaction between the charged mediator (with charge $Q_1$) and the charged body (e.g. a nearby fixed mirror with charge $Q_2$). The Coulomb interaction can be realized experimentally in an optomechanical system. In experiment \cite{Rabl2010,Arcizet2011}, an ensemble of NV centers and electric charges can be embedded in a single crystal diamond nanobeam. Alternatively, a charged particle can also be levitated in a magnetic trap or Paul trap \cite{Rudolph2022,Rieser2022,Slezak2019,Kuhlicke2014,DelordNjp2017,Serafini2009}. Since the mechanical excitations (or phonons) occur in pair, this is also referred to as an effective two-phonon drive. With the two-phonon drive, the effective interaction between the TP and the qubit is \textit{exponentially} enhanced compared with the case without the two-phonon drive. The protocol differs from the known proposals that probe for gravitationally induced entanglement in at least three points: (i) the mediator does not need a large mass due to the exponential enhancement of the gravitational interaction, (ii) we need to increase the two-phonon driving strength only slightly to significantly accelerate the generation of entanglement in certain regions, and (iii) this entanglement can be partially preserved in presence of dephasing by the proposed strategy. These advantages bring experiments on gravity-induced entanglement a step closer to reality.

The remainder of this manuscript is organized as follows. In Sec.~\ref{sec2}, a physical model is introduced to describe the setup, and the dynamical equations for the system are derived. In Sec.~\ref{sec3}, we study the entanglement between the two quantum systems. Sec. ~\ref{sec4} devotes to the entanglement analysis in the presence of decoherence. In Sec. ~\ref{sec5}, we discuss the experimental feasibility of the protocol. Finally,  we conclude  in Sec.~\ref{sec6}. In the Appendix \ref{appendix}, we discuss the validity of the approximation done in Eq. (\ref{Hamiltonian4}).

\begin{figure}[h]
\centering
\includegraphics[width=0.48\textwidth]{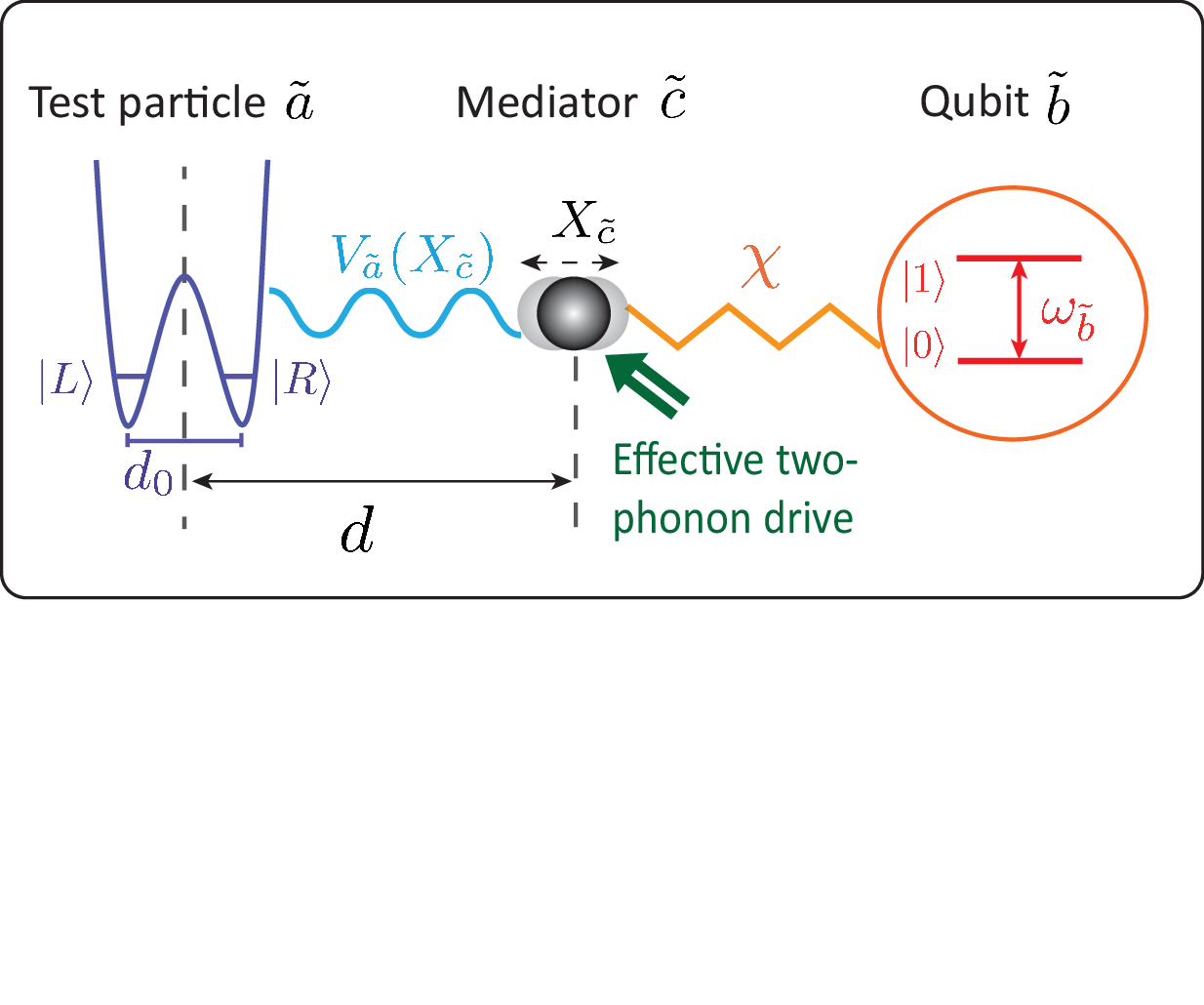}\vspace{-2.5cm}
\caption {(Color online) Schematic diagram of the hybrid quantum device for probing the quantum nature of gravity. A test particle (system $\tilde{a}$) is trapped in a double-well potential (DWP) forming  state $|A\rangle$ ($A=L,R$) and the distance between the two wells is denoted by $d_0$. A qubit (system $\tilde{b}$)  that has transition frequency $\omega_{\tilde{b}}$, is coupled to a mechanical oscillator (system $\tilde{c}$) with magnetic coupling strength $\chi$. Levels $\left|0\right\rangle$ and $\left|1\right\rangle$ correspond to the ground state and the excited state of the qubit, respectively. The mechanical oscillator is introduced to mediate the interaction between systems $\tilde{a}$ and $\tilde{b}$. The equilibrium position of the mediator is located at a distance $d$ from the center of the DWP. $X_{\tilde{c}}$ is the displacement of the mediator from equilibrium. The mediator is weakly coupled to the TP through gravitational interaction with energy $V_{\tilde{a}}(X_{\tilde{c}})$. The direct coupling between systems $\tilde{a}$ and $\tilde{b}$ is ignored. To achieve a effective two-phonon drive, the mediator is coupled to a nearby fixed mirror via the Coulomb interaction. See the text for details.}
	\label{model}

\end{figure}

\section{Modeling and dynamics}\label{sec2}
Consider the general situation illustrated in Fig \ref{model}. The system $\tilde{a}$ is a TP of mass $m_{\tilde{a}}$ trapped in a DWP along one spatial direction (e.g., the $X$ direction) and behaves as a two-level system with frequency $\omega_{\tilde{a},0}$ and state $|A\rangle$ ($A=L,R$). We assume that the wells are far and deep enough to make any tunneling term negligible, and thus, that states $|L\rangle$ and $|R\rangle$ are stationary states of the DWP for the duration of the protocol. The system $\tilde{b}$ is a two-level atom (qubit) with frequency $\omega_{\tilde{b}}$ and state $|B\rangle$ ($B=0,1$). Finally, the system $\tilde{c}$ is a mechanical oscillator (mediator) with mass $m_{\tilde{c}}$ and frequency $\omega_{\tilde{c}}$, and consider its motion in only one spatial direction, e.g., the $X$ direction. Assuming that the mediator is weakly coupled to the TP via gravitational interaction with a interaction energy $V_{\tilde{a}}(X_{\tilde{c}})$. To ensure that the gravitational energy dominates over Casimir forces \cite{Emig2007}, the separation distance between the interacting objects needs to be sufficiently large, typically exceeding significantly the separation $d_0$ of the DWP \cite{Plenio2022}. We further take into account the mediator couples to the qubit via the direct spin-phonon magnetic coupling \cite{Lee2017} and interacts with a nearby fixed mirror through the Coulomb interaction \cite{Ekinci2005,Tian2004,Hensinger2005,Zhang2012,Bochmann2013,Bekker2017,Feng2021Two,Penny03123,
Rieser2022,Rudolph2022,Serafini2009}. The Hamiltonian that describes  such a system reads,
\begin{equation}
\begin{aligned}
H=H_f+H_i+H_{co}.
\label{Hamiltonian0}
\end{aligned}
\end{equation}
Here the free Hamiltonian $H_f$ takes,
\begin{equation}
\begin{aligned}
H_f= \sum_{A=L, R} \hbar \omega_{\tilde{a},0} |A\rangle\langle A|+\hbar \omega_{\tilde{b}} \sigma_{\tilde{b}}^{z}+\frac{P_{\tilde{c}}^{2}}{2 m_{\tilde{c}}}+\frac{1}{2} m_{\tilde{c}} \omega_{\tilde{c}}^{2} X_{\tilde{c}}^{2},
\end{aligned}
\end{equation}
the interaction Hamiltonian $H_i$ represents the TP-mediator and qubit-mediator couplings,
\begin{equation}
\begin{aligned}
H_i=\sum_{A=L, R} V_{\tilde{a}, A} (X_{\tilde{c}})|A\rangle\langle A|
+\hbar \chi X_{\tilde{c}} \sigma_{\tilde{b}}^{z},
\end{aligned}
\end{equation}
and the Coulomb interaction is
\begin{equation}
\begin{aligned}
H_{co}=\frac{k_e Q_1 Q_2}{\left|r_0+X_{\tilde{c}}\right|}.
\end{aligned}
\end{equation}
Here $\omega_{\tilde{a},0}=0$ due to the zero energy splitting of the TP. $\sigma_{\tilde{b}}^{z}=|1\rangle\langle 1|-| 0\rangle\langle 0|$ is the Pauli $z$ operator acting on the system $\tilde{b}$ with the transition frequency $\omega_{\tilde{b}}$. $X_{\tilde{c}}$ and $P_{\tilde{c}}$ are the position and momentum operators of the mediator, respectively. The term $V_{\tilde{a}} (X_{\tilde{c}})$ represents the interaction energy between systems $\tilde{c}$ and $\tilde{a}$. The coefficient $\chi$ is introduced for the qubit-mediator coupling strength, which is independent of the mass of the qubit \cite{Lee2017}. The direct coupling between systems $\tilde{a}$ and $\tilde{b}$ is negligible. In the Coulomb interaction Hamiltonian $H_{co}$, $k_e$ is the electrostatic constant. $Q_1>0$ is the positive charge on the charged mediator, and $Q_2<0$ is the negative charge on the charged body. $r_0$ denotes the equilibrium separation between the charged mediator and the charged body, in absence of any interaction between them.

By expanding the Newtonian gravitational energy in the limit $\pm d_0/2-X_{\tilde{c}}\ll d$, one gets \cite{Qvarfort2019}
\begin{widetext}
\begin{equation}
\begin{aligned}
V_{\tilde{a}, \pm}= & -\frac{G m_{\tilde{a}} m_{\tilde{c}}}{\left|d \mp \frac{d_0}{2}+X_{\tilde{c}}\right|}=-\frac{G m_{\tilde{a}} m_{\tilde{c}}}{d}\left(1+\frac{d_0^2}{4 d^2} \pm \frac{d_0}{2 d}-\left(1 \pm \frac{d_0}{d}\right) \frac{X_{\tilde{c}}}{d}+\frac{X_{\tilde{c}}^2}{d^2}+\cdots\right),
\label{expansion1}
\end{aligned}
\end{equation}
\end{widetext}
where $G$ is the Newton constant, and $V_{\tilde{a}, +}$ and $V_{\tilde{a}, -}$ correspond to $V_{\tilde{a}, R}$ and $V_{\tilde{a}, L}$, respectively. The first two terms in the expansion produce the static energy shift, the third term gives the energy splitting of the TP, while the fourth term generates the displacement of the oscillator equilibrium position and the linear interaction between the TP and the mediator. Finally, the fifth term introduces a shift in the oscillation frequency of the mediator. As the term proportional to $(\pm \frac{d_0}{2}-X_{\tilde{c}})^2$ is already very small under usual
experimental conditions, we ignore all terms of order $\mathcal{O}$$\left[\left(\pm \frac{d_0}{2}-X_{\tilde{c}}\right)^3\right]$. Similarly, in the case of $X_{\tilde{c}}\ll r_0$, $H_{co}$ can be expanded as \cite{Tian2004,Hensinger2005,Feng2021Two}
\begin{eqnarray}
\begin{aligned}
H_{co}=-\frac{k_e \left|Q_1\right| \left|Q_2\right|}{r_0}\left(1-\frac{X_{\tilde{c}}}{r_0}+\frac{X_{\tilde{c}}^2}{r_0^2}+\cdots\right),
\label{expansion2}
\end{aligned}
\end{eqnarray}
where we ignore all terms of order $\mathcal{O}$$\left(X_{\tilde{c}}^3\right)$.  The validity of this approximation is manifested in Sec. ~\ref{sec5}.

Introducing position and momentum operators for the mediator as $X_{\tilde{c}}=\left[\hbar /\left(2 m_{\tilde{c}} \omega_{\tilde{c}}\right)\right]^{1 / 2}\left(a^{\dagger}+a\right)$ and $P_{\tilde{c}}=i\left[\hbar m_{\tilde{c}} \omega_{\tilde{c}} / 2\right]^{1 / 2}\left(a^{\dagger}-a\right)$,
the above Hamiltonian of the system can be rewritten as
\begin{eqnarray}
\begin{aligned}
\tilde{H}=&\hbar \omega_{\tilde{a}} \sigma_{\tilde{a}}^{z}+\hbar \omega_{\tilde{b}} \sigma_{\tilde{b}}^{z}+\hbar \left(\tilde{\omega}-2F\right) a^{\dagger} a+\hbar\epsilon\left(a+a^{\dagger}\right)\\
&-\hbar F\left(a^{\dagger 2}+a^2\right)+\hbar\left(g_{\tilde{a}} \sigma_{\tilde{a}}^{z}+g_{\tilde{b}} \sigma_{\tilde{b}}^{z}\right)\left(a+a^{\dagger}\right),
\label{Hamiltonian4}
\end{aligned}
\end{eqnarray}
where we defined $\sigma_{\tilde{a}}^{z}=|L\rangle\langle L|-| R\rangle\langle R|$ and the energy splitting of the system $\tilde{a}$ is given by $\omega_{\tilde{a}}=\omega_{\tilde{a},0}+G m_{\tilde{a}} m_{\tilde{c}} d_{0} /\left(2 \hbar d^{2}\right)$. $\tilde{\omega}^{2}=\omega_{\tilde{c}}^{2}-\left(2 G m_{\tilde{a}} / d^{3}\right)$ is the modified frequency of the mediator, $F=k_e \left|Q_1\right| \left|Q_2\right| /\left(2 m_{\tilde{c}} \omega_{\tilde{c}} r_0^3\right)$ is the effective two-phonon driving strength, and the fourth term comes from the linear term proportional to $X_{\tilde{c}}$ in the expansion of the gravitational and Coulomb interactions with $\epsilon=\left(G m_{\tilde{a}} m_{\tilde{c}}/d^2+k_e \left|Q_1\right| \left|Q_2\right| /r_0^2 \right)\left[1/\left(2 \hbar m_{\tilde{c}} \omega_{\tilde{c}}\right)\right]^{1 / 2}$. Since this displacement term does not affect the entanglement \cite{Serafini2017}, we can safely ignore the linear term in position, which can be verified by numerical simulations (more details are given in the Appendix \ref{appendix}). The TP-mediator and qubit-mediator coupling strengths are $g_{\tilde{a}}=-\left(G m_{\tilde{a}} d_{0} / d^{3}\right) \sqrt{m_{\tilde{c}} /(2 \tilde{\omega} \hbar)}$ and $g_{\tilde{b}}=\chi \left[\hbar /\left(2 m_{\tilde{c}} \omega_{\tilde{c}}\right)\right]^{1 / 2}$, respectively. Note that the above two-phonon drive can also be realized for the case where the spring constant of the mechanical resonator is modulated via applying a tunable and time-varying voltage to this electrode \cite{Woolley2008,Szorkovszky2011,Szorkovszky2013,Szorkovszky2014,Clerk2016,Nori2020}. In contrast, this scheme of Coulomb interaction-enhanced entanglement is much simpler and does not require additional modulation of the spring constant of the mechanical oscillator. It is worth mentioning that the mechanical resonator electrometers usually do not measure the charge densities in tiny objects (e.g., $<$6 nm) \cite{Lammine2011}. In Eq. (\ref{Hamiltonian4}), we have ignored the zero-point energy from the third term.
\begin{figure}[h]
	\centering

	\includegraphics[width=0.5\textwidth]{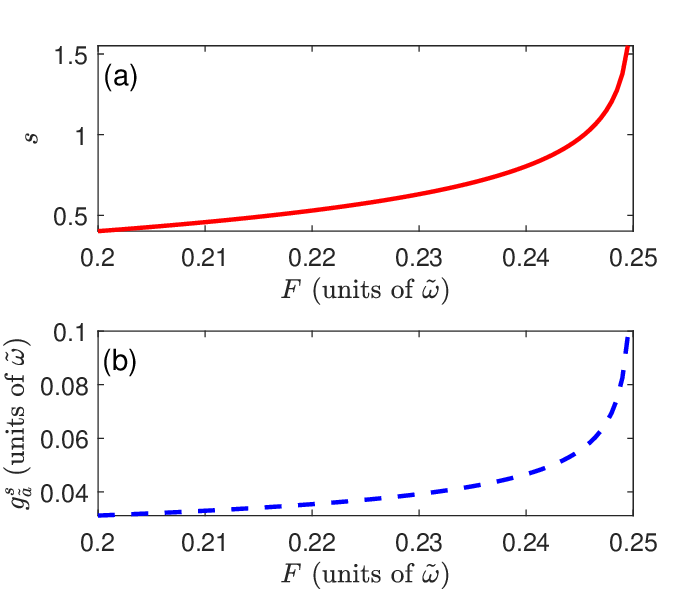}
	\caption {(Color online) (a) The squeezing parameter $s$ versus two-phonon driving strength $F$. (b) The gravitational interaction strength $g_{\tilde{a}}^s$ versus the two-phonon driving strength $F$. The parameters used are chosen as $g_{\tilde{a}}=1/48\tilde{\omega}$ and $g_{\tilde{b}}=\tilde{\omega}$, as in Ref.  \cite{Plenio2022}.}
	\label{Fig2}

\end{figure}

In order to obtain strong TP-mediator and qubit-mediator coupling strengths, we need to use the mechanical amplification of mediator. By virtue of a Bogoliubov squeezing transformation \cite{Bogoljubov1958} $a=\cosh (s)a_s +\sinh (s)a_s^{\dagger}$, with a squeezing parameter $s=(1 / 4) \ln \left[\tilde{\omega} /\left(\tilde{\omega}-4 F\right)\right]$ and a preferred squeezed mechanical mode. Then, we can obtain the Hamiltonian for this hybrid system in this squeezed frame
\begin{eqnarray}
\begin{aligned}
H_s=&\hbar \omega_{\tilde{a}} \sigma_{\tilde{a}}^{z}+\hbar \omega_{\tilde{b}} \sigma_{\tilde{b}}^{z}+\hbar \omega_s a_s^{\dagger} a_s\\
&+\hbar\left(g^s_{\tilde{a}} \sigma_{\tilde{a}}^{z}+g^s_{\tilde{b}} \sigma_{\tilde{b}}^{z}\right)\left(a_s+a_s^{\dagger}\right).
\label{Hamiltonian5}
\end{aligned}
\end{eqnarray}
\begin{figure}[h]
	\centering

	\includegraphics[width=0.5\textwidth]{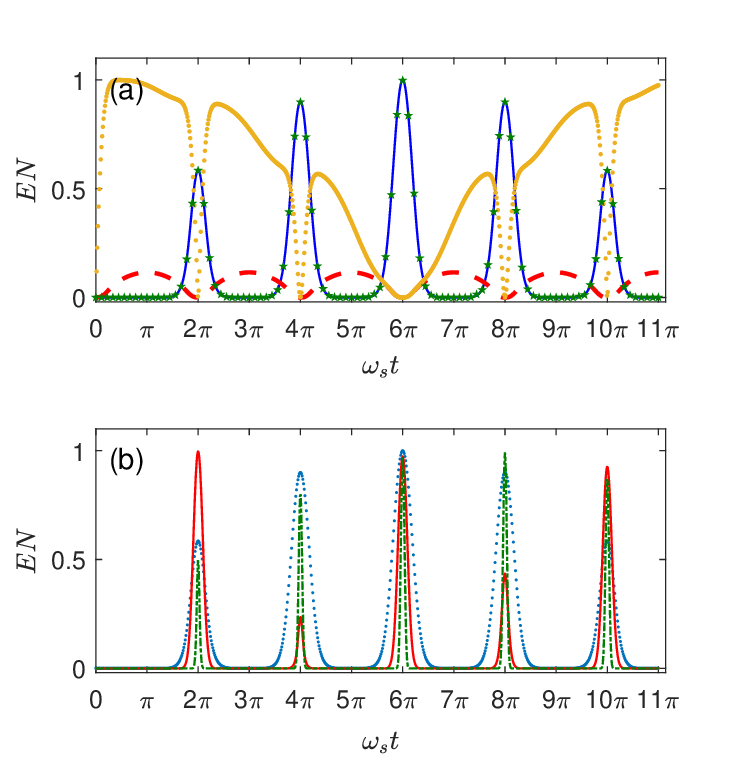}
	\caption {(Color online) Temporal evolution of  the entanglement  quantified
by the logarithmic negativity $EN$. (a) The blue solid line denotes the $EN$ between the TP and the qubit, the red dashed line represents the $EN$ between the TP and the mediator, and the orange solid circles display the $EN$ between the qubit and the mediator. The green solid stars are the numerical results obtained using the Hamiltonian (\ref{Hamiltonian5}), which is independent of $\omega_{\tilde{a}}$ and $\omega_{\tilde{b}}$. (b) The evolution of entanglement between the TP and the qubit at three different squeezing parameter. We vary $\delta=\tilde{\omega}$ ($s=0$, blue solid circles), $0.5\tilde{\omega}$ ($s=0.1733$, red solid line), and $0.2\tilde{\omega}$ ($s=0.4024$, green dash-dotted line). Here, the symbol $\delta$ represents the difference between parameters $\tilde{\omega}$ and $4F$ as $\delta=\tilde{\omega}-4F$. The other system parameters are the same as in Fig. \ref{Fig2}. These curves are plotted in the absence of a decoherence process.}
	\label{LN_1}
\end{figure}
Here, the third and fifth terms in Eq. (\ref{Hamiltonian4}) have been diagonalized by the squeezing transformation and are simplified to a squeezed mediator $(\hbar \omega_s a_s^{\dagger} a_s)$ in Eq. (\ref{Hamiltonian5}) with a transformed mechanical frequency $\omega_s=\sqrt{\tilde{\omega}(\tilde{\omega}-4F)}=(\tilde{\omega}-4F)\exp(2s)$. In the context, the local squeezing enhances the coupling strength as
\begin{equation}
\begin{aligned}
g^s_{j}=g_je^s (j=\tilde{a},\tilde{b}),
\end{aligned}
\end{equation}
where the TP-mediator and qubit-mediator coupling strengths grow \textit{exponentially} with the squeezing parameter $s$. Such an exponential
enhancement of this interaction is one of our most important results. We notice that the effective squeezed mode frequency $\omega_s$, the squeezing parameter $s$, and the coupling strength $g^s_j$ are controlled by the modified frequency $\tilde{\omega}$ of the mediator and the two-phonon driving strength $F$, as shown in Fig. \ref{Fig2}. When $\tilde{\omega}$ infinitely approaches $4F$, the coupling strength $g^s_j$ is enhanced significantly with respect to the bare coupling strength $g_j$. We would demonstrate later how the introduction of the squeezing parameter promptly enhances and accelerates the generation of entanglement between the TP and the qubit, with the entanglement is highly sensitive to variations in the squeezing parameter.

In a rotating reference frame defined by $U=\exp\left(-i\omega_{\tilde{a}} \sigma_{\tilde{a}}^{z}t-i\omega_{\tilde{b}} \sigma_{\tilde{b}}^{z}t-i\omega_s a_s^\dagger a_s t\right)$, the Hamiltonian (\ref{Hamiltonian5}) is reduced to
\begin{equation}
H_{I}(t)=\left(g^s_{\tilde{a}} \sigma_{\tilde{a}}^{z}+g^s_{\tilde{b}} \sigma_{\tilde{b}}^{z}\right)\left(a_s e^{-i \omega_s t}+a_s^{\dagger} e^{i \omega_s t}\right).
\label{HI}
\end{equation}
Notice that Eq. (\ref{HI}) has a communtation relation at different times given by
\begin{small}
\begin{equation}
\left[H_{I}(t), H_{I}\left(t^{\prime}\right)\right]=\left(g^s_{\tilde{a}} \sigma_{\tilde{a}}^{z}+g^s_{\tilde{b}} \sigma_{\tilde{b}}^{z}\right)^{2}\left(e^{i \omega_s\left(t^{\prime}-t\right)}-e^{-i\omega_s \left(t^{\prime}-t\right)}\right),
\end{equation}
\end{small}
and all higher-order communtation relation vanish. Thus, the  time evolution in the interaction picture is characterized by the first two orders in the Magnus expansion $U_{I}=\exp \left\{\Theta_1(t)+\Theta_2(t)\right\}$ \cite{Magnus1954}, with
\begin{eqnarray}
\begin{aligned}
\Theta_1(t)&=-i \int_{0}^{t} d t^{\prime} H\left(t^{\prime}\right)=-i\left(g^s_{\tilde{a}} \sigma_{\tilde{a}}^{z}+g^s_{\tilde{b}} \sigma_{\tilde{b}}^{z}\right)
\\&\left(a \frac{e^{-i \omega_s t}-1}{-i \omega_s}+a^{\dagger} \frac{e^{i \omega_s t}-1}{i \omega_s}\right), \\
\Theta_2(t)&=-\frac{1}{2} \int_{0}^{t} d t^{\prime} \int_{0}^{t^{\prime}} d t^{\prime \prime}\left[H\left(t^{\prime}\right), H\left(t^{\prime \prime}\right)\right]\\&=i\left(g^s_{\tilde{a}} \sigma_{\tilde{a}}^{z}+g^s_{\tilde{b}} \sigma_{\tilde{b}}^{z}\right)^{2}\left(t / \omega_s-\frac{\sin \omega_s t}{\omega_s^{2}}\right) .
\end{aligned}
\end{eqnarray}

The time evolution of the system is then performed with the
unitary operator
\begin{equation}
\begin{aligned}
U_I=& \exp \left\{\left(g^s_{\tilde{a}} \sigma_{\tilde{a}}^{z}+g^s_{\tilde{b}} \sigma_{\tilde{b}}^{z}\right)\left(-a_s \alpha_{t}+a_s^{\dagger} \alpha_{t}^{*}\right)\right\} \\
& \times \exp \left\{-i \frac{2 g^s_{\tilde{a}} g^s_{\tilde{b}}}{\omega_s} \sigma_{\tilde{a}}^{z} \sigma_{\tilde{b}}^{z}\left(t-\frac{\sin \omega_s t}{\omega_s}\right)\right\},
\label{unitary operator}
\end{aligned}
\end{equation}
where we defined $\alpha_{t}=\left[\left(e^{-i \omega_s t}-1\right) / \omega_s\right]$. At times $t_n=2\pi n/\omega_s$ that are a natural period of the mediator, then $\alpha_{t}=0$  and the first term vanishes,
leaving an effective interaction between the TP and the
qubit with an effective coupling $\tilde{g}_{\text{eff}}=2 g^s_{\tilde{a}} g^s_{\tilde{b}}/\omega_s$. Then the unitary operator Eq. (\ref{unitary operator}) is reduced to
\begin{equation}
U_I\left(t_{n}\right)=\exp \left\{-i \tilde{g}_{\text{eff}} \sigma_{\tilde{a}}^{z} \sigma_{\tilde{b}}^{z} t_{n}\right\}.
\end{equation}
Therefore, at these points in time the mediator is correlated and decorrelated periodically from the rest of the system, and the initial mediator state does not impact the quantum entanglement between the TP and the qubit. This entanglement can then be detected by standard approaches making local measurements on the 2-qubit system \cite{Plenio2007,Horodecki2009}. Note, however, that decoherence ensuing from damping to the mediator motion during the state evolution will adversely affect the final detection of the entanglement. We will not consider this phonon decoherence in this work, and instead assume that the mechanical decoherence is
negligible over one oscillation period of the mediator \cite{Mancini1997}.

In the following, we will introduce the concept of a logarithmic negativity $EN$ and clarify that the role of the two-phonon drive for enhancing and accelerating the generation of entanglement between the TP and the qubit.

\section{entanglement}\label{sec3}
In this section, we consider that the TP and the qubit are prepared in the states $(|L\rangle+|R\rangle) / \sqrt{2}$ and $(|1\rangle+|0\rangle) / \sqrt{2}$ , respectively. For the sake of convenience, we assume that the mediator is initially in a squeezed coherent state $|\zeta\rangle=S(\xi)|\alpha\rangle$ with the
squeezing operator $S(\xi)=\exp \left(\frac{1}{2} \xi^{*} a_s^{2}-\frac{1}{2} \xi a_s^{\dagger 2}\right)$ and the
squeezing parameter $\xi=s \exp (\mathrm{i} \theta)$ ($\theta=\pi$). Therefore, the initial state is given by $|\psi_{\text {in}}\rangle=\frac{1}{2}\sum_{A,B} \left | A ; \text {in}\right \rangle_{\tilde{a}} \left | B ; \text {in}\right \rangle_{\tilde{b}} |\zeta;\text {in}\rangle_{\tilde{c}}$ $(A=L,R;B=0,1)$.

After a time $t$ in the considered system, the evolved state of the system is given by
\begin{widetext}
\begin{eqnarray}
\begin{aligned}
|\psi_{\text {out}}(t)\rangle=&U_f
U_I(t)|\psi_{\text {in}}\rangle\\
=&\frac{1}{2}U_f\Big\{e^{[-i(\varphi-\Phi_1)]}|R,0\rangle |\alpha_1,\zeta\rangle+e^{[-i(\varphi-\Phi_2)]}|L,1\rangle |\alpha_2,\zeta\rangle +e^{[i(\varphi+\Phi_3)]}|R,1\rangle |\alpha_3,\zeta\rangle+e^{[i(\varphi+\Phi_4)]}|L,0\rangle |\alpha_4,\zeta\rangle \Big\},
\end{aligned}
\end{eqnarray}
\end{widetext}
where
\begin{eqnarray}
\begin{aligned}
U_f&=e^{-i (\hbar \omega_{\tilde{a}} \sigma_{\tilde{a}}^{z}+\hbar \omega_{\tilde{b}} \sigma_{\tilde{b}}^{z}+\hbar \omega_s a_s^{\dagger} a_s) t/\hbar},\\
\varphi&=\frac{2g^s_{\tilde{a}} g^s_{\tilde{b}}}{\omega_s}\left(t-\frac{\sin \omega_s t}{\omega_s}\right),\\
\Phi_{1}&=\Phi_{2}=\text{Im}[(g^s_{\tilde{a}}\alpha_t^*)(g^s_{\tilde{b}} \alpha_t^*)^*], \\
\Phi_{3}&=\Phi_{4}=\text{Im}[(-g^s_{\tilde{a}}\alpha_t^*)(g^s_{\tilde{b}} \alpha_t^*)^*],\\
\alpha_1&=-(g^s_{\tilde{a}}+g^s_{\tilde{b}})\alpha_t^*,\quad
\alpha_2=(g^s_{\tilde{a}}+g^s_{\tilde{b}})\alpha_t^*,\\
\alpha_3&=-(g^s_{\tilde{a}}-g^s_{\tilde{b}})\alpha_t^*,\quad
\alpha_4=(g^s_{\tilde{a}}-g^s_{\tilde{b}})\alpha_t^*.
\end{aligned}
\end{eqnarray}

To quantify the entanglement between the TP and the qubit, we introduce the positive partial transpose (PPT) criterion \cite{Peres1996,Horodecki1996} and the logarithmic negativity $EN$ \cite{Plenio2007} as follows. For a given density operator $\rho_{\tilde{a}\tilde{b}}$ of systems $\tilde{a}$ and $\tilde{b}$, the partial transposition matrix $\rho_{\tilde{a}\tilde{b}}^{T_{\tilde{b}}}$ of $\rho_{\tilde{a}\tilde{b}}(t)=\text {Tr}_{\tilde{c}}\left[|\psi_{\text {out}}(t)\rangle \langle\psi_{\text {out}}(t)|\right]$ is given by
\begin{equation}
{ }_{\tilde{a}}\left\langle\left. A^{\prime}\right|_{\tilde{b}}\left\langle B^{\prime}\left|\rho_{\tilde{a}\tilde{b}}^{\mathrm{T}_{\tilde{b}}}\right| A\right\rangle_{\tilde{a}} \mid B \right\rangle_{\tilde{b}}={ }_{\tilde{a}}\left\langle\left. A^{\prime}\right|_{\tilde{b}}\langle B|\rho_{\tilde{a}\tilde{b}}| A\rangle_{\tilde{a}} \mid B^{\prime}\right\rangle_{\tilde{b}}
\end{equation}
for a basis $\left\{|A\rangle_{\tilde{a}}|B\rangle_{\tilde{b}}\right\}$ of the Hilbert space $\mathcal{H}_{\tilde{a}} \otimes \mathcal{H}_{\tilde{b}}$ of systems $\tilde{a}$ and $\tilde{b}$. $T_{\tilde{b}}$ denotes the partial transpose with respect to the subsystem $\tilde{b}$. If $\rho_{\tilde{a}\tilde{b}}^{T_{\tilde{b}}}$
has a negative eigenvalue then the density matrix $\rho_{\tilde{a}\tilde{b}}$ is entangled. This is called the PPT criterion \cite{Peres1996,Horodecki1996}. We obtain an entanglement measure based on the PPT criterion as
\begin{equation}
E N=\max \left(0, \log _{2}\left\|\rho_{\tilde{a}\tilde{b}}^{T_{\tilde{b}}}\right\|_{1}\right),
\label{EN}
\end{equation}
where $\left\|\cdot\right\|_{1}$ is the trace norm, which is defined as $\|\chi\|_{1}=\operatorname{Tr} \sqrt{\chi \chi^{\dagger}}$, and $EN\in[0,1]$ with the maximum value corresponding to maximally entangled states and the minimum value to separable states.

The partial transpose matrix $\rho_{\tilde{a}\tilde{b}}^{T_{\tilde{b}}}$ takes the following form:
\begin{small}
\begin{widetext}
\begin{equation}
\frac{1}{4}\left(
 \begin{array}{cccc}
 1&e^{[i(2\varphi+\Phi_3-\Phi_1)]} \langle\alpha_1,\zeta \mid \alpha_3,\zeta\rangle&e^{[i(-2\varphi+\Phi_1-\Phi_4)]} \langle\alpha_4,\zeta \mid \alpha_1,\zeta\rangle&e^{[i(\Phi_3-\Phi_4)]} \langle\alpha_4,\zeta \mid \alpha_3,\zeta\rangle\\
 *&1&e^{[i(\Phi_1-\Phi_2)]} \langle\alpha_2,\zeta \mid \alpha_1,\zeta\rangle&e^{[i(2\varphi+\Phi_3-\Phi_2)]} \langle\alpha_2,\zeta \mid \alpha_3,\zeta\rangle\\
 *&*&1&e^{[i(-2\varphi+\Phi_2-\Phi_4)]}\langle\alpha_4,\zeta \mid \alpha_2,\zeta\rangle\\
 *&*&*&1
 \end{array}
\right),
\label{partial transpose}
\end{equation}
\end{widetext}
\end{small}
where $*$s are determined by the Hermiticity of $\rho_{\tilde{a}\tilde{b}}^{T_{\tilde{b}}}$. The four rows and four columns are labeled by the out states $\left|R,0;\text {out}\right \rangle, \left|R,1;\text {out}\right \rangle, \left|L,0 ;\text{out}\right \rangle, \left|L,1;\text {out}\right \rangle$ and $\left\langle R,0;\text {out} \right |, \left\langle R,1;\text {out} \right |, \left\langle L,0;\text {out} \right |, \left\langle L,1;\text {out} \right |$, respectively. The out states
\begin{eqnarray}
\begin{aligned}
\left | A ; \text { out }\right \rangle_{\tilde{a}}&=\exp\left (-i \omega_{\tilde{a}} \sigma_{\tilde{a}}^{z} t\right)\left | A ; \text {in}\right \rangle_{\tilde{a}},\\
\left|B ; \text { out } \right \rangle_{\tilde{b}}&=\exp\left(-i \omega_{\tilde{b}} \sigma_{\tilde{b}}^{z} t\right)\left | B ; \text {in}\right \rangle_{\tilde{b}}
\label{out states}
\end{aligned}
\end{eqnarray}
describe the states of the TP and the qubit at the time $t$, respectively \cite{Matsumura2022}. In the local Hamiltonians $\hbar \omega_{\tilde{a}} \sigma_{\tilde{a}}^{z}$ and $\hbar \omega_{\tilde{b}} \sigma_{\tilde{b}}^{z}$ may give other accumulated phases, which are included in the out states $\left | A ; \text { out }\right \rangle_{\tilde{a}}$ and
$\left | B ; \text { out }\right \rangle_{\tilde{b}}$. Since such phases are given by the local unitary evolution, they do not influence the entanglement between the TP and the qubit, which can be verified by numerical simulations [see the green solid stars in Fig. \ref{LN_1}(a)].
\begin{figure}[h]
	\centering

	\includegraphics[width=0.5\textwidth]{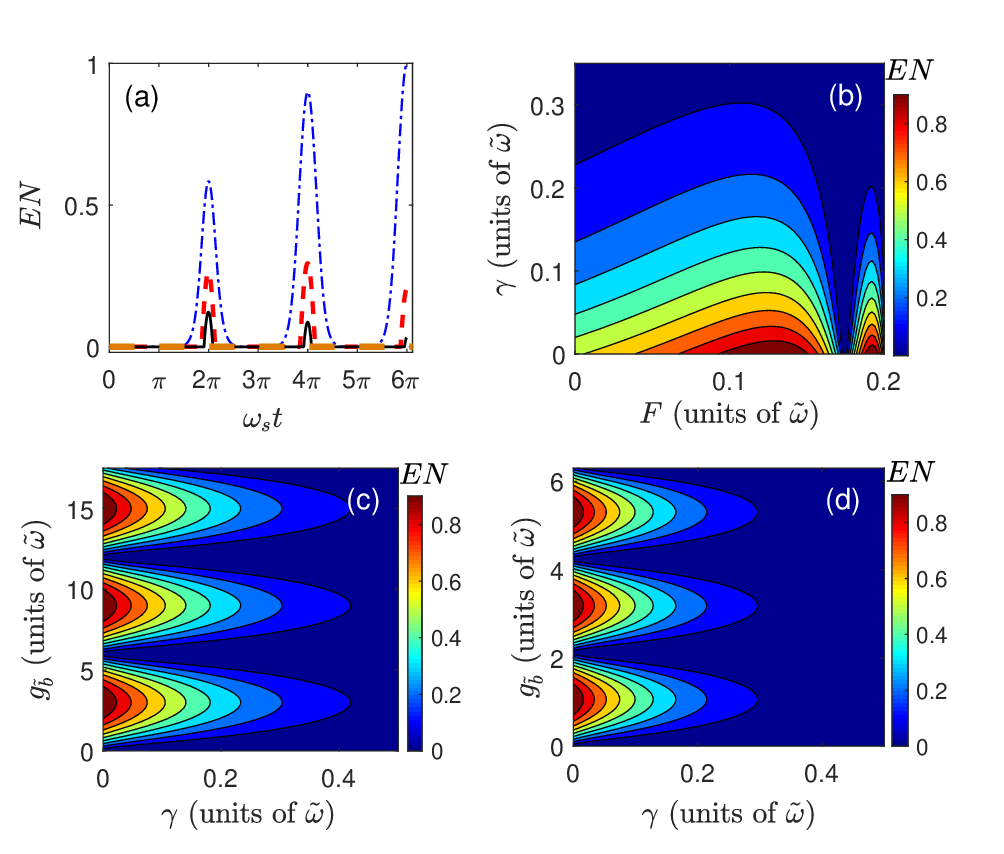}
	\caption {(Color online) The behavior of the $EN$ of systems $\tilde{a}$ and $\tilde{b}$. (a) $EN$ $vs$ $t$ at $F=0$ ($s=0$). The blue dash-dotted line shows the $EN$ at $\gamma=0$, the thin red dashed line indicates the $EN$ at $\gamma=0.1\tilde{\omega}$, the black solid line denotes the $EN$ at $\gamma=0.2\tilde{\omega}$, and the thick orange dashed line represents the $EN$ at $\gamma=0.33\tilde{\omega}$. (b) $EN$ $vs$ $F$ and $\gamma$ at $\omega_s t_n=2\pi$. (c) $EN$ $vs$ $g_{\tilde{b}}$ and $\gamma$ at $\omega_s t_n=2\pi$ and $\delta=\tilde{\omega}$ ($s=0$, $\omega_s=\tilde{\omega}$). (d) $EN$ $vs$ $g_{\tilde{b}}$ and $\gamma$ at $\omega_s t_n=2\pi$ and $\delta=0.5\tilde{\omega}$ ($s=0.1733$, $\omega_s=0.7071\tilde{\omega}$). The other system parameters are the same as in Fig. \ref{Fig2}.}
	\label{LN_2}

\end{figure}

As shown in Fig. \ref{LN_1}(a), we see that the entanglement (red dashed line) between the TP and the mediator oscillates with the period of the mediator and disappears completely at times $t_n$, as predicted by Eq. (\ref{unitary operator}). A similar result was obtained for the entanglement (orange solid circles) between the qubit and the mediator. At these times the mediator is decoupled while the logarithmic negativity between the TP and the qubit reaches its maximum (blue solid line). The maximum entanglement is given by
\begin{equation}
E N(t_n)=\max \big\{0, \log _2[1+|\sin (2 \tilde{g}_{\text{eff}} t_n)|]\big\}.
\end{equation}
For comparison, we numerically solve the logarithmic negativity between the TP and the qubit by the Hamiltonian (\ref{Hamiltonian5}) [see the green solid stars]. We see that the above result (blue solid line) is highly consistent with the numerical result. In Fig. \ref{LN_1}(b), we show that the entanglement between the TP and the qubit can be enhanced via introducing the two-phonon drive. It can be observed that the amount of generated entanglement does not increase monotonically with the squeezing parameter.

\section{decoherence}\label{sec4}
Until now, we implicitly considered the unitary dynamics imprinting on the quantum systems. However, quantum entanglement is fragile in the sense that when the
quantum system is in contact with an environment \cite{Eberly2002,Eberly2003,Budini00400,Matsumura2022Path,Matsumura2022}. In this
section, we consider how the decoherence influences the entanglement generation between the TP and the qubit. A simple case arises if the decoherence process is assumed to affect only one system, say qubit. We consider the usual situation where the qubit energy relaxation time is much longer than the dephasing time \cite{Reina2002,Schreier2008,Clerk2008,Delord2017}. Such suppression can be modeled by introducing the decaying factor as
\begin{equation}
\begin{aligned}
&\mid \mathrm{1} ; \text { in }\rangle_{\tilde{b}}\left\langle\mathrm{0} ; \text { in }\left|\rightarrow e^{-\gamma t}\right| \mathrm{1} ; \text { out }\right\rangle_{\tilde{b}}\langle\mathrm{0} ; \text { out }|, \\
&\mid \mathrm{0} ; \text { in }\rangle_{\tilde{b}}\left\langle\mathrm{1} ; \text { in }\left|\rightarrow e^{-\gamma t}\right| \mathrm{0} ; \text { out }\right\rangle_{\tilde{b}}\langle\mathrm{1} ; \text { out }|,
\end{aligned}
\end{equation}
where the decaying term $e^{-\gamma t}$ fully characterizes the dephasing dynamics of the qubit. In this analysis, the dephasing rate is treated as a phenomenological parameter. Taking into account the dephasing rate, the modified partial transpose matrix is obtained as follows

\begin{equation}
\begin{aligned}
\frac{1}{4}\left(
 \begin{array}{cccc}
 1&\rho_{1,2}&\rho_{1,3}&\rho_{1,4}\\
 \rho_{1,2}^*&1&\rho_{2,3}&\rho_{2,4}\\
 \rho_{1,3}^*&\rho_{2,3}^*&1&\rho_{3,4}\\
 \rho_{1,4}^*&\rho_{2,4}^*&\rho_{3,4}^*&1
 \end{array}
\right),
\label{decoherence}
\end{aligned}
\end{equation}
whose elements are given by
\begin{equation}
\begin{aligned}
\rho_{1,2}=&e^{[i(2\varphi+\Phi_3-\Phi_1)-\gamma t]} \langle\alpha_1,\zeta \mid \alpha_3,\zeta\rangle,\\
\rho_{1,3}=&e^{[i(-2\varphi+\Phi_1-\Phi_4)]} \langle\alpha_4,\zeta \mid \alpha_1,\zeta\rangle,\\
\rho_{1,4}=&e^{[i(\Phi_3-\Phi_4)-\gamma t]} \langle\alpha_4,\zeta \mid \alpha_3,\zeta\rangle,\\
\rho_{2,3}=&e^{[i(\Phi_1-\Phi_2)-\gamma t]} \langle\alpha_2,\zeta \mid \alpha_1,\zeta\rangle,\\
\rho_{2,4}=&e^{[i(2\varphi+\Phi_3-\Phi_2)]} \langle\alpha_2,\zeta \mid \alpha_3,\zeta\rangle,\\
\rho_{3,4}=&e^{[i(-2\varphi+\Phi_2-\Phi_4)-\gamma t]}\langle\alpha_4,\zeta \mid \alpha_2,\zeta\rangle.
\end{aligned}
\end{equation}
Similarly, the effect of TP dephasing can also be considered.
At times $t_n=2\pi n/\omega_s$, then $\alpha_j= 0$, $\Phi_j=0$, and $\varphi=\frac{4\pi n g^s_{\tilde{a}} g^s_{\tilde{b}}}{\omega_s^2}$ ($j=1,2,3,4$). Thus, the final state Eq. (\ref{decoherence}) is reduced to
\begin{small}
\begin{equation}
\frac{1}{4}\left(
 \begin{array}{cccc}
 1&e^{i2\varphi-\gamma t_n} &e^{-i2\varphi} &e^{-\gamma t_n} \\
 e^{-i2\varphi-\gamma t_n} &1&e^{-\gamma t_n} &e^{i2\varphi} \\
 e^{i2\varphi}&e^{-\gamma t_n}&1&e^{-i2\varphi-\gamma t_n}\\
 e^{-\gamma t_n}&e^{-i2\varphi}&e^{i2\varphi-\gamma t_n}&1
 \end{array}
\right).
\label{decoherence2}
\end{equation}
\end{small}

\begin{figure}[h]
	\centering

	\includegraphics[width=0.5\textwidth]{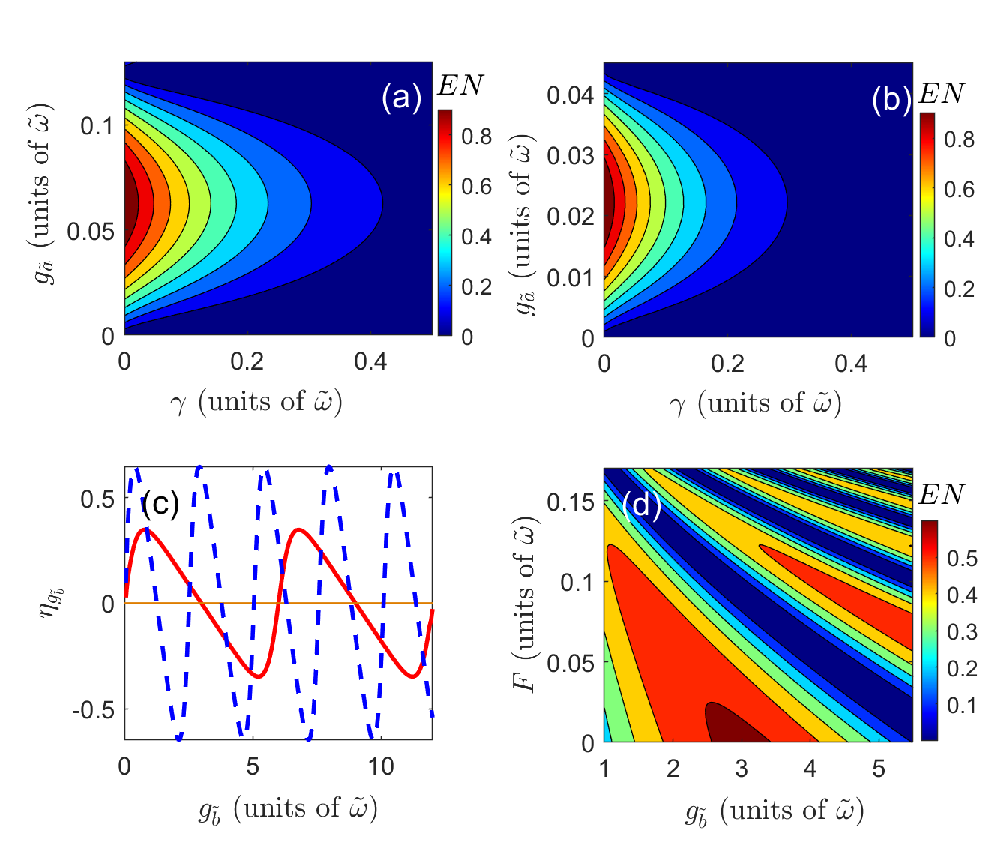}
	\caption {(Color online) (a) $EN$ $vs$ $g_{\tilde{a}}$ and $\gamma$ at $\omega_s t_n=2\pi$, $g_{\tilde{b}}=\tilde{\omega}$, and $\delta=\tilde{\omega}$ ($s=0$, $\omega_s=\tilde{\omega}$). (b) $EN$ $vs$ $g_{\tilde{a}}$ and $\gamma$ at $\omega_s t_n=2\pi$, $g_{\tilde{b}}=\tilde{\omega}$, and $\delta=0.5\tilde{\omega}$ ($s=0.1733$, $\omega_s=0.7071\tilde{\omega}$). (c) The generation rate $\eta_{g_{\tilde{b}}}$ of entanglement between the TP and the qubit at $\gamma =0.1\tilde{\omega}$, $\omega_s t_n=2\pi$, and $g_{\tilde{a}}=1/48\tilde{\omega}$. We vary $s=0$ (red solid line) and $s=0.1733$ (blue dashed line). The orange solid line represents that the rate of entanglement generation is zero. (d) $EN$ $vs$ $F$ and $g_{\tilde{b}}$ at $\omega_st_n=2\pi$, $\gamma =0.1\tilde{\omega}$, and $g_{\tilde{a}}=1/48\tilde{\omega}$. The squeezing parameter satisfies $s\in[0,0.2849]$. }
	\label{Fig5}

\end{figure}

In Fig. \ref{LN_2}(a), we show the evolution of the entanglement between the TP and the qubit, but now in the presence of dephasing rate of the qubit. Without mechanical amplification, the entanglement between the TP and the qubit is seriously spoiled by the detrimental decoherence. However, the amount of generated entanglement can be enhanced to some extent by increasing the two-phonon driving strength [see Fig. \ref{LN_2}(b)]. To amplify the role of the two-phonon drive, we plot the logarithmic negativity $EN$ between the TP and the qubit in ($\gamma$, $g_{\tilde{b}}$) parameter's space [see Figs. \ref{LN_2}(c) and \ref{LN_2}(d)]. We observe that the entanglement exists only within a certain range of $\gamma$. More importantly, the rate of entanglement generation after the introduction of the two-phonon drive [see Fig. \ref{LN_2}(d)] is much faster than the rate before the introduction of the two-phonon drive [see Fig. \ref{LN_2}(c)]. Similarly, we plot the logarithmic negativity $EN$ between the TP and the qubit in ($\gamma$, $g_{\tilde{a}}$) parameter's space [see Figs. \ref{Fig5}(a) and \ref{Fig5}(b)]. It can be seen that the introduction of the two-phonon drive greatly reduces the requirement on the mass of the mediator.

To quantify the generation rate of entanglement between the TP and the qubit, we define the rate of generated entanglement as
\begin{eqnarray}
\begin{aligned}
\eta_{g_{j}}=\frac{d(EN)}{d g_{j}} \quad (j=\tilde{a},\tilde{b}).
\end{aligned}
\end{eqnarray}
Figure \ref{Fig5}(c) shows $\eta_{g_{\tilde{b}}}$ as a function of the coupling strength $g_{\tilde{b}}$. The orange solid line represents that the generation rate of entanglement is kept at zero. Above this solid line, the rate of entanglement generation speeds up and, vice-versa, slows down. Moreover, we need to increase the two-phonon driving strength only slightly to considerably accelerate the generation of entanglement in certain regions, which is another major finding of our analysis.

Considering the amount of entanglement between the TP and the qubit does not increase monotonically with the squeezing parameter. To this end, we discuss the possibility of enhancing the amount of entanglement between the TP and the qubit by jointly modulating the squeezing parameter $s$ and the qubit-mediator coupling strength $g_{\tilde{b}}$ [see Fig. \ref{Fig5}(d)]. Specifically, when the squeezing parameter is modulated to a value corresponding to a minimum amount of entanglement, the amount of entanglement between the TP and the qubit can be further improved by modulating the coupling strength $g_{\tilde{b}}$.

\section{Discussion of experimental feasibility}\label{sec5}
To examine the feasibility of this proposal in the experiments, we consider a particle of nanodiamond \cite{aaa} in a DWP as a TP with radii $r_{\tilde{a}}=70$ nm, density $\rho_{\tilde{a}}=3.5\times10^3$ Kg/m$^3$ (mass $m_{\tilde{a}}=5.0286\times 10^{-18}$ Kg) and separation $d_0=500$ nm. Recently, the trapping and cooling of nanodiamonds have been demonstrated with magnetic traps \cite{Hs2016,Twamley2019}. There are many types of mechanical oscillators \cite{Rudolph2022,Arcizet2011,Rieser2022,Serafini2009,LeeNjp2011,Cirac2012,Kuhlicke2014,
DelordNjp2017,Slezak2018,Slezak2019,Plenio2020,Nori2020,Delord2017,ZhangAdv2021}. For concreteness, we select one of them, where the mechanical degree of freedom (mediator) is a charged diamond particle containing a single nitrogen vacancy (NV) center. The diamond particle is levitated in a magnetic trap or Paul trap and cooled down to an internal temperature of 1 K. It possesses a density of $\rho_{\tilde{c}}=3.5\times10^3$ Kg/m$^3$, radii $r_{\tilde{c}}=1.25$ $\mu$m (mass $m_{\tilde{c}}=2.8634\times 10^{-14}$ Kg), and an oscillation frequency $\omega_{\tilde{c}}/2\pi=2$ KHz ($\omega_{\tilde{c}}\simeq\tilde{\omega}$) \cite{Kuhlicke2014}. The equilibrium position of the mediator is located at a separation distance $d=180$ $\mu$m from the center of the DWP. Based on the above parameters, we find that the gravitational interaction strength is $g_{\tilde{a}}=2.9907\times 10^{-15}$ Hz. To magnetically couple a single NV center (qubit) to the mediator, the NV spin must be immersed in a spatially inhomogeneous magnetic field such that the position of the NV spin in the field is controlled by the position of the mediator \cite{Lee2017}. The possible first-order gradient magnetic field $B^{\prime}$ in the laboratory is about $10^{4}$ T$/$m \cite{Slezak2018}. We can obtain the magnetic coupling strength between the mediator and the qubit as $g_{\tilde{b}}=\gamma_e B^{\prime} \left[\hbar /\left(2 m_{\tilde{c}} \omega_{\tilde{c}}\right)\right]^{1 / 2} = 673.4614$ Hz, where $\gamma_e/2\pi= 28$ GHz/T is the electronic gyromagnetic ratio.

It is worth noting that the decoherence process for the NV spin should be considered. The single NV spin decoherence in diamond is mainly induced by the coupling of the surrounding electron or nuclear spins \cite{Nori2020,Plenio2020,Delord2017}. With the advancement of the dynamical decoupling techniques \cite{Walsworth2013,Du2009Preserving,Naydenov2011,Zhao2012Decoherence,
Liu2008Universality,Ryan2010}, the dephasing time for a single NV center in diamond is estimated to be $1/\gamma \simeq 1$ ms. We find that the resulting entanglement between TP and qubit vanishes at $\omega_s t=2\pi$ and large decoherence. To generate and enhance the entanglement between the TP and the qubit, it is necessary to enhance the effective TP-qubit coupling strength $\tilde{g}_{\text{eff}}$. However, until now this experimental implementation of such gravitationally mediated entanglement is still a big challenge. Fortunately, in our proposal, this effective interaction can be dramatically enhanced by exploiting the two-phonon drive. In order
to introduce a two-phonon drive, we consider an ultra-miniature charged body can be embedded in a nearby fixed mirror \cite{Rabl2010,Arcizet2011}. The relevant parameters of the Coulomb interaction can be chosen as $\left|Q_1\right|=10^{-15}$ C and $\left|Q_2\right|=27.5$ nC \cite{Kuhlicke2014,Zhang2012}, and then the two-phonon driving strength $F$ can be estimated via $r_0$ ($\sim10^{-3}$ m). When $\tilde{\omega}$ infinitely approaches $4F$, we obtain a large squeezing parameter $s$. This leads to the realization of the strong effective interaction $\tilde{g}_{\text{eff}}$. We assume that the difference between parameters $\tilde{\omega}$ and $4F$ as $\delta=\tilde{\omega}-4F=10^{-8}$ Hz, the transformed mechanical frequency is given by $\omega_s=\sqrt{\tilde{\omega}(\tilde{\omega}-4F)}=0.0112$ Hz and the squeezing parameter is $s=(1 / 4) \ln \left[\tilde{\omega} /\left(\tilde{\omega}-4 F\right)\right]=6.9649$. Then, we find that the TP-mediator and qubit-mediator coupling strengths are $g^s_{\tilde{a}}=3.1665\times 10^{-12}$ Hz and $g^s_{\tilde{b}}=7.1304\times10^{5}$ Hz, respectively ($\tilde{g}_{\text{eff}}=2 g^s_{\tilde{a}} g^s_{\tilde{b}}/\omega_s= 4.0283\times 10^{-4}$ Hz). Based on the above parameters, our protocol theoretically predicts that the amount of entanglement between TP and qubit falls in the range of $EN\in[0.5224,0.001]$ at $\omega_s t=2\pi$ and $\gamma\in[0,0.01]$ Hz. However, this dephasing rate is beyond what can be attained with the state-of-the-art instruments and requires further experimental efforts.

On the other hand, the present scheme works for zero dephasing rate of TP. In realistic scenarios, however, the disturbance of TP by environmental noise is inevitable. The electrostatic interaction (if the TP carries a net charge, it will be subject to a random Coulomb interaction due to fluctuating electric fields), the magnetic interaction due to fluctuating magnetic field, and the direct magnetic dipole interaction between the TP and the qubit are potential sources of noise. It is fortunate that there are mitigation methodologies for this electromagnetic interactions \cite{Monteiro2020,Frimmer2017,Bose2017}. Moreover, we note that in the presence of the mechanical amplification, the mechanical noise is also amplified. To eliminate the adverse effect of amplified mechanical noises, a possible approach is to use the dissipative squeezing method (see the supplementary of Ref. \cite{Nori2020}), where an additional optical mode is added to the system for suppressing the mechanical noise.

Finally, we check the validity of the approximation done in Eq. (\ref{expansion2}). Note that the center-of-mass (COM) delocalization of the mechanical oscillator $\Delta_x$ is given by $\Delta_x\simeq \sqrt{\hbar/m_{\tilde{c}} \tilde{\omega}}\exp(s)$.
Based on the above parameters, the COM delocalization is $5.7318\times 10^{-10}$ m. At this point, the third term in Eq. (\ref{expansion2}) is extremely small, so we can safely ignore all terms of order $\mathcal{O}$$\left(X_{\tilde{c}}^3\right)$. The interaction between TP and mediator, dominated by Casimir force, remains negligible (the minimum distance between their surfaces is kept above 157 $\mu$m \cite{Plenio2022}).

Although it is extremely challenging to attain the required experimental parameters, at the least, the present work provides a different approach for experimental detection of the quantum nature of gravity.

\begin{figure}[h]
	\centering

	\includegraphics[width=0.5\textwidth]{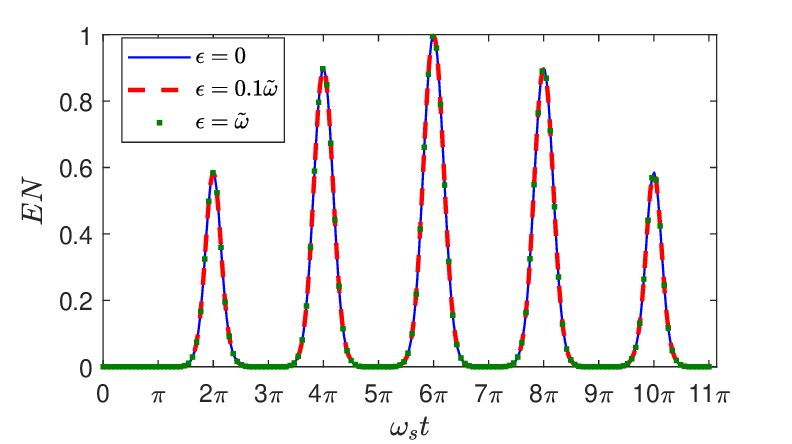}
	\caption {(Color online) Temporal evolution of the degree of the entanglement, as quantified by the logarithmic negativity $EN$. Here we assume an
initial state $(|L\rangle+|R\rangle) / \sqrt{2} \otimes(|1\rangle+|0\rangle) / \sqrt{2} \otimes |\alpha\rangle$, where $(|L\rangle+|R\rangle) / \sqrt{2}$ is the initial state of the TP, $(|1\rangle+|0\rangle) / \sqrt{2}$ is the initial state of the qubit, and $|\alpha\rangle$ is the initial coherent state of the mediator. We vary $\epsilon=0$ (blue solid line), $0.1\tilde{\omega}$ (red dashed line), and $\tilde{\omega}$ (green solid squares). The parameters used are chosen as $g_{\tilde{a}}=1/48\tilde{\omega}$, $g_{\tilde{b}}=\tilde{\omega}$, and $F=0$. These curves are plotted in the absence of a decoherence process.}
	\label{Fig6}
\end{figure}

\section{Conclusion}\label{sec6}
In conclusion, we have proposed a mechanism to enhance the gravity-induced entanglement via the use of two-phonon drive. The system consists of a test particle (TP), a qubit, and a quntum mediator. Among the three parts, the TP is coupled to the mediator via a gravitational interaction, while the qubit is coupled to the mediator through a direct spin-phonon coupling. A two-phonon drive is introduced  via the Coulomb interaction between the charged mediator and the charged body. The introduction of the two-phonon drive allows us to amplify the effective interaction between the TP and the qubit that grows exponentially with the squeezing parameter. More interestingly, the generation of entanglement between the TP and the qubit can be significantly improved and accelerated compared with the case without the two-phonon drive. Our protocol can partially recover the entanglement between the TP and the qubit in presence of dephasing. It is worth noting that the basic mechanism proposed in this manuscript is generic, i.e., it is not restricted to the specific case under discussion but can be extended to other defect centers or solid-state systems such as germanium-vacancy center, silicon-vacancy center, and
tin-vacancy center in diamond \cite{Hepp2014,Bhaskar2017,Iwasaki2017,Thiering2018}.

  \section{acknowledgments}\label{sec7}
This work is supported by National Natural Science Foundation of China (NSFC) under Grants No. $12175033$ and No. $12147206$  and National Key R$\&$D Program of China under Grant No. 2021YFE0193500.

\begin{appendix}

\section{The validity of the approximation done in the main text}\label{appendix}
In this Appendix, we discuss the validity of the approximation done in Eq. (\ref{Hamiltonian4}). In Fig. \ref{Fig6}, we numerically plot the temporal evolution of the degree of the entanglement between the TP and the qubit by the Hamiltonian (\ref{Hamiltonian4}) and Eq. (\ref{EN}). Clearly, the displacement term in the Hamiltonian (\ref{Hamiltonian4}) does not affect the entanglement between the TP and the qubit, as we expected.

\end{appendix}


\begin{thebibliography}{99}
\bibitem{aaa}We assume that the diamond is a perfect sphere with density $\rho = 3.5$ g/cm$^3$.
\end{thebibliography}
\end{document}